\documentclass{svproc}
\usepackage{lipsum}
\usepackage{url}
\usepackage[hidelinks]{hyperref}
\usepackage{graphicx}
\usepackage{caption}
\usepackage{subcaption}
\usepackage{xcolor}
\usepackage{soul}
\usepackage{siunitx}
\usepackage{multirow}
\usepackage[numbers]{natbib}
\usepackage{amsmath, amssymb}
\usepackage{mathtools}
\usepackage{cases}
\usepackage{bm}
\usepackage{enumerate}
\usepackage{algorithm}
\usepackage{algpseudocode}
\usepackage{accents}
\usepackage{booktabs}
\usepackage{tablefootnote}
\newcommand{\tridx}[1]{^{\{#1\}}}

\begin{document}
\mainmatter 

\title{Accounts of using the Tustin-Net architecture on a rotary inverted pendulum}

\titlerunning{Accounts of using the Tustin-Net architecture on a rotary inverted pendulum}

\author{Stijn van Esch\inst{1} \and Fabio Bonassi\inst{2}
 \and Thomas B. Sch\"on\inst{2}}

\authorrunning{Stijn van Esch et al.}

\tocauthor{Stijn van Esch, Fabio Bonassi, and Thomas B. Sch\"on\inst{2}}

\institute{Eindhoven University of Technology, Eindhoven, The Netherlands,\\
\email{s.j.a.v.esch@student.tue.nl},
\and
Uppsala University, Department of Information Technology,\\
Uppsala, Sweden, \\ \email{\{fabio.bonassi, thomas.schon\}@it.uu.se}}

\maketitle          

\begin{abstract}
In this report we investigate the use of the Tustin neural network architecture (Tustin-Net) for the identification of a physical rotary inverse pendulum.
This physics-based architecture is of particular interest as it builds on the known relationship between velocities and positions.
We here aim at discussing the advantages, limitations and performance of Tustin-Nets compared to first-principles grey-box  models on a real physical apparatus, showing how, with a standard training procedure, the former can hardly achieve the same accuracy as the latter.
To address this limitation, we present a training strategy based on transfer learning that yields Tustin-Nets that are competitive with the first-principles model, without requiring extensive knowledge of the setup as the latter.
\end{abstract}

\section{Introduction} \label{sec:intro}
Modeling a dynamic system is a notoriously difficult, knowledge-intensive, and time-consuming task.
In the realm of mechanical systems, there are well-established techniques for modeling nonlinear dynamical systems, such as the Euler-Lagrange equations \cite{cheli2015advanced}, but they of course require a thorough knowledge of the physical system at hand and domain expertise.
Often, the resulting models are nonetheless grey-box, i.e., they depend on unknown parameters or functions that needs to be identified from the data.

With the advent of deep learning researchers have thus begun to investigate the possibility of leveraging the universal approximation capabilities of Neural Networks (NNs) to learn such models directly from data \cite{masri1993identification}, minimizing the system knowledge required to retrieve an accurate model.
Recently, gated recurrent neural networks, such as Long Short-Term Memory networks and Gated Recurrent Units, have demonstrated  appealing performance for nonlinear system identification  \cite{bonassi2022recurrent, bonassi2023deep, pillonetto2023deep,  schoukens2019nonlinear} and control \cite{bonassi2022imc, bonassi2024nonlinear}.
These architectures, however, are deemed unsuitable for identifying mechanical systems because they call for the data to be sampled from a system with  stability and boundedness properties. 
If not, the resulting models may be accurate only over short simulation horizons~\cite{sangiorgio2020robustness}.

One promising solution for this identification task are continuous-time Neural ODEs \cite{rahman2022neural}.
These architectures feature a feedforward NN as nonlinear state function, and they are trained by first computing the state and output trajectories by means of an ODE solver, and then minimizing the simulation error via automatic differentiation tools powering modern deep learning libraries like PyTorch~\cite{paszke2019pytorch}.

Following the trend of physics-based machine learning \cite{willard2020integrating}, it has been recently noted that imposing a physically-consistent structure on the state function yields models that are easier to train, with better performance and enhanced consistency to the physical laws governing the system.
In \cite{pozzoli2020tustin}, this strategy is embraced to build the \textbf{Tustin-Net} architecture.
These discrete-time neural models are characterized by the fact that  the dynamics of the positions is fixed to the integral---discretized by trapezoidal (Tustin) method---of the velocities, while the discrete-time velocities dynamics are governed by a feedforward NN approximating the velocities' increments. 

\medskip
The goal of this report is to discuss the application of Tustin-Nets to a Quanser rotary inverse pendulum apparatus, validating the performance of this architecture on a real-world system and comparing it with a first-principles model obtained via the Euler-Lagrange equations.
This architecture, to the best of our knowledge, has not yet been tested on a real apparatus.
In addition, we propose a training procedure based on transfer learning \cite{tan2018survey} to improve the performance of Tustin-Nets even in the presence of unbalanced datasets, in which there are large intervals where the system is at equilibrium.
This procedure consists of pre-training the Tustin-Net on a dataset extracted from the initial transients of the experiments that constitute the training data, after which the weights of the final layers of the network are fine-tuned on the entire dataset to enhance their performance.

\medskip
The report is structured as follows. In Section \ref{sec:mechanical}, the physical setup is presented and a first-principles model is derived via the Euler-Lagrange equations. 
The Tustin-Net architecture is then introduced in Section~\ref{sec:tustin-net}, and the transfer learning-based training procedure is presented. 
The numerical results are reported in Section~\ref{sec:numerical}, showing that such a procedure allows Tustin-Nets to achieve enhanced accuracy. 
Conclusions are eventually drawn in Section~\ref{sec:conclusions}.

\subsection*{Notation}
Given a real vector $v \in \mathbb{R}^n$, we let $v^\prime$ be its transpose and $\| v \|_2$ its $2$-norm. 
Moreover, its $j$-th component is denoted as $[v]_j$.
For a vector $v(t)$ dependent on the continuous-time $t$, the time dependency is generally omitted for compactness. 
Vectors depending on the discrete-time index $k$ are instead denoted as $v_k$. A sequence between the time-steps $k_1$ and $k_2$ is defined as
$v_{k_1:k_2} = \{ v_{k_1}, v_{k_1 + 1}, ...,  v_{k_2} \}$, and we let $\textrm{cat}(v_{k_1:k_2}) = [ v_{k_1}^\prime, v_{k_1+1}^\prime, ..., v_{k_2}^\prime ]^\prime$.

At last, given a function $L(x_1, x_2): \mathbb{R}^n \times \mathbb{R}^n \to \mathbb{R}$, we let $\nabla_{x_i} L(x_1, x_2)$ be its gradient with respect to $x_i$.

\section{First-principles model} \label{sec:mechanical}
In this section, the rotary pendulum apparatus used for the experiments is described, and the first-principles grey-box model derived from the Euler-Lagrange equations is introduced.

\subsection{Setup}
\begin{figure}[t]
	\centering
     \begin{subfigure}[b]{0.4 \linewidth}
         \centering
         \includegraphics[height=0.28\textheight]{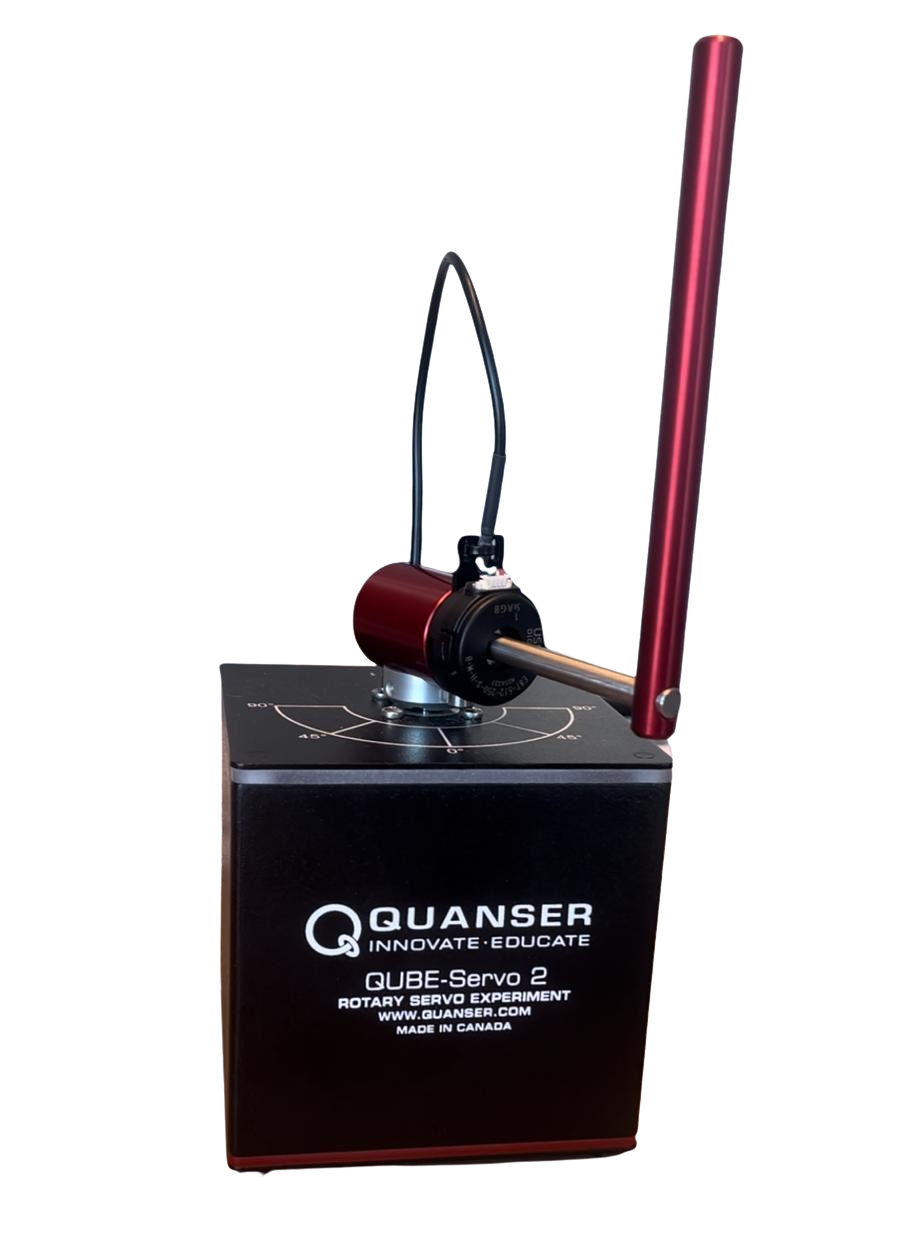}
     \end{subfigure}
     \hfill
     \begin{subfigure}[b]{0.5 \linewidth}
         \centering
         \includegraphics[height=0.28\textheight]{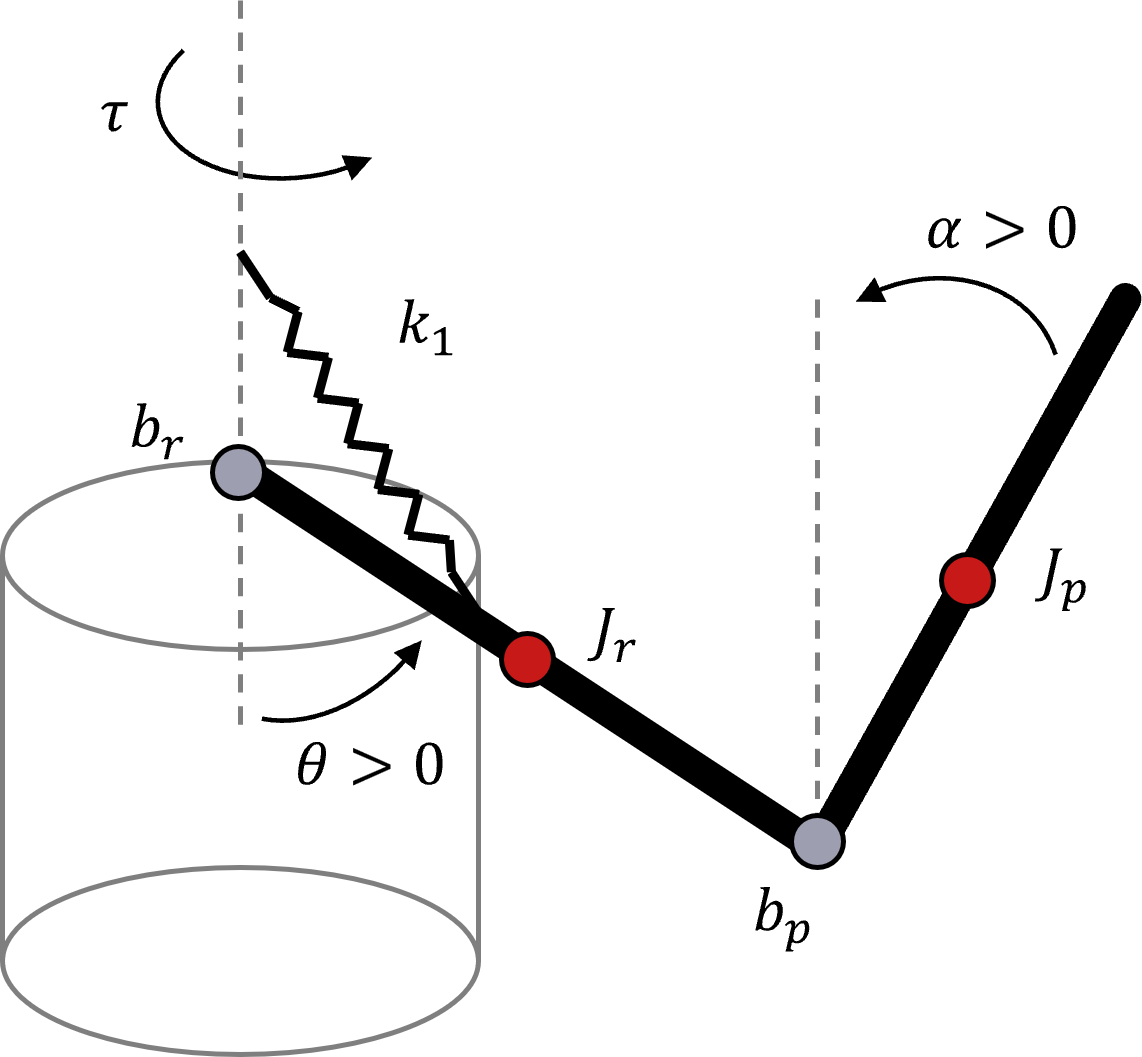}
     \end{subfigure}
     \caption{Rotary pendulum system: picture of the lab apparatus (left) and schematic overview (right).}
     \label{fig:pendulum}
\end{figure}

In this work, we consider the problem of modeling the rotational pendulum shown in \autoref{fig:pendulum}  along with its schematic diagram.
The apparatus is a Quanser Qube Servo 2 system \cite{quanserqube2} in its rotary pendulum configuration.

The system consists of a rotational arm, which forms an angle $\theta$ with respect to the center line, and a pendulum, which forms an angle $\alpha$ with respect to the vertical axis.
These angles are assumed positive in the counterclockwise direction with respect to the arms' axes of rotation, and they are measured by encoders with resolution $\frac{\pi}{1024}$ rad per count in quadrature \cite{quanserqube2}.

The system is equipped with a DC motor that applies a torque $\tau$ to the base arm that depends on the input voltage $U$ applied to the motor itself.
This voltage is of course subject to saturation, $U \in [ - 15 \textrm{V}, 15 \textrm{V}]$.

It is important to point out that the encoder of the pendulum arm is connected to the logical board of the apparatus by an external cable.
As discussed in Section~\ref{sec:mechanical:el}, this cable introduces a nonlinear elastic force on the base arm, which needs to be accounted in order to build an accurate physical model of the rotary pendulum.

\subsection{Euler-Lagrange grey-box model} \label{sec:mechanical:el}
One way to retrieve a first-principles model of the rotary pendulum is to apply the Euler-Lagrange equations, that allow to derive a dynamical model from the expression of the energy and non-conservative forces acting on the system.
To this end, we let $q = [ \theta, \alpha ]^\prime$ be the vector of generalized coordinates of the system, and we note that the system can be fully described by $q$ and the vector of velocities  $\dot{q} = [ \dot{\theta}, \dot{\alpha}]^\prime$. 
The Euler-Lagrange equations of the system read
\begin{equation}\label{eq:eulerlagrange}
	\frac{d}{dt} \nabla_{\dot{q}} L(q, \dot{q}) - \nabla_{q} L(q, \dot{q}) = Q_{nc}(U, \dot{q}),
\end{equation}
where $Q_{nc}$ denotes the non-conservative forces and $L(q, \dot{q})$ denotes the Lagrangian function, i.e.,
\begin{subequations}\label{eq:lagrangian}
\begin{equation} \label{eq:lagrangian:lagrangian_def}
    L(q, \dot{q}) = T(q, \dot{q}) - V(q),
\end{equation}
with $T(q, \dot{q})$ and $V(q)$ being the kinetic and  potential energy of the system, respectively.  
For the system at hand, the kinetic energy can be compactly expressed as
\begin{equation}
	T(q, \dot{q}) = \frac{1}{2} \dot{q}^\prime J(q; \Theta) \dot{q},
\end{equation}
where $J(q; \Theta)$ is the matrix of generalized inertia.
This expression is omitted here for compactness, but we highlight that $J$ depends on some not easily measurable parameters, such as motor inertia, which are collected in the set $\Theta$. 
These parameters will need to be identified from the data, as discussed later this section.
Letting $z_r(q)$ and $z_p(q)$ denote the z-axis coordinate of the center of masses of the two arms with respect to a reference frame fixed at the center of the motor, the potential energy is then given by
\begin{equation} \label{eq:lagrangian:potential}
	V(q) = m_r g z_r(q) + m_p g z_p(q) + V_s(q; \Theta).
\end{equation}
In this expression, the first two terms are the gravitational potential energy, while the last term is a polynomial elastic potential energy that models the spring introduced by the encoder's cable,
\begin{equation}  \label{eq:lagrangian:spring}
    V_s(q; \Theta) = \frac{1}{2}\kappa_1 \theta^2 + \frac{1}{3} \kappa_2 \theta^3,
\end{equation}
where the coefficients $\kappa_1$ and $\kappa_2$ belong to the set of parameters to identify, $\Theta$.

At last, since the system is non-conservative, the term $Q_{nc}$ is introduced to capture the dissipative forces and the effect of the actuator. In particular
\begin{equation}  \label{eq:lagrangian:Qnc}
    Q_{nc}(U, \dot{q}; \Theta) = \begin{bmatrix}
        1 \\ 0
    \end{bmatrix} \tau(U, \dot{q}; \Theta) -  \begin{bmatrix}
        b_1 \\ b_2
    \end{bmatrix} \dot{q},
\end{equation}
where $\tau(U, \dot{q}; \Theta) = \kappa_t\, (U - \kappa_v\dot{\theta})$ denotes the torque applied by the DC motor and the second term is a linear damping term.
Note that the parameters $\kappa_v$, $\kappa_t$, $b_1$, and $b_2$  need to be identified, and are hence part of $\Theta$.
\end{subequations}

\medskip
Substituting the resulting Lagrangian \eqref{eq:lagrangian} into the Euler-Lagrange equations~\eqref{eq:eulerlagrange} results in
\begin{equation} \label{eq:mechanical:explicit_eulerlagrange}
	M(q; \Theta) \ddot{q} + C(q, \dot{q}; \Theta) \dot{q} + H(q; \Theta) = K(\Theta) U,
\end{equation} 
where $M(q; \Theta)$ is invertible since $J(q; \Theta)$ does not have singularities.
The grey-box state space model can hence be formulated as
\begin{subnumcases}{\label{eq:mechanical:accelerations}}
	\frac{d}{dt} q = \dot{q}, \label{eq:mechanical:accelerations:1}\\
	\frac{d}{dt} \dot{q} = - [M(q; \Theta)]^{-1} \big( C(q, \dot{q}; \Theta) \dot{q} + H(q; \Theta)\big) + [M(q; \Theta)]^{-1} K(\Theta) u. \label{eq:mechanical:accelerations:2}
\end{subnumcases}
Letting $x = [ q^\prime, \dot{q}^\prime ]^\prime$ and $y = q$, we will compactly denote the grey-box Euler-Lagrange model as 
\begin{equation} \label{eq:mechanical:statespace}
\begin{dcases}
	\frac{d}{dt} x = \varphi(x, u; \Theta) \\
	y = \begin{bmatrix}
		I & 0
	\end{bmatrix} x
\end{dcases}
\end{equation}

\subsubsection{Parameters identification} \label{sec:mechanical:training}
The grey-box model \eqref{eq:mechanical:statespace} depends upon the set of parameter $\Theta$. 
Because these parameters cannot be easily measured, they are identified from the \emph{training data}, which consists of $N$ tuples $\big(x^{\{i\}}_{0}, u^{\{i\}}_{0:T_i}, y^{\{i\}}_{0:T_i}\big)$ indexed by $i \in \{1, ..., N \}$, where the input and output sequences are sampled with a sampling time $\tau_s$. 
The training set thus reads
\begin{equation} \label{eq:mechanical:dataset}
	\mathcal{D} = \left\{ \big(x_0\tridx{i}, u_{0:T_i}\tridx{i}, y\tridx{i}_{0:T_i}\big), \forall i \in \{1, ..., N \} \right\}.
\end{equation}
The identification is carried out by minimizing the \emph{loss function}
\begin{equation} \label{eq:mechanical:loss}
	\mathcal{L}_q(\mathcal{D}; \Theta) = \frac{1}{\lvert \mathcal{D} \lvert} \sum_{i\in \mathcal{D}} \frac{1}{T_i} \sum_{k=1}^{T_i} \sum_{j=1}^{n_y} \ell_j\Big(\big[y_k(x_0\tridx{i}, u_{0:k}\tridx{i}; \Theta)\big]_j, \big[y_{k}\tridx{i}\big]_j \Big),
\end{equation}
where $\ell_j$ is a distance measure between the $j$-th component of the sampled output vector, i.e. $\big[y_{k}\tridx{i}\big]_j$, and the simulated output denoted by $y_k(x_0\tridx{i}, u_{0:k}\tridx{i}; \Theta)$.
Note that this latter is obtained by numerically solving  the ODE \eqref{eq:mechanical:statespace} with initial conditions $x_0\tridx{i}$ and input signal $u_{0:k}\tridx{i}$ at instant $t = k \tau_s$ for $k \in \{ 0, ...,  T_i \}$.
We here take 
\begin{equation} \label{eq:mechanical:distance}
    \ell_j([\hat{y}_k]_j, [{y}_k]_j) = \big([\hat{y}_k]_j - [{y}_k]_j \big)^2,
\end{equation}
so that \eqref{eq:mechanical:loss} represents the mean squared simulation error of \eqref{eq:mechanical:statespace} over the training dataset $\mathcal{D}$. 
The parameters are therefore estimated by solving 
\begin{equation} \label{eq:mechanical:identification}
\begin{aligned}
	\Theta^\star = \arg\min_\Theta \,\, & \mathcal{L}_q(\mathcal{D}; \Theta) \\
	 \text{s.t.} \quad & \Theta \in \Xi.
\end{aligned}	
\end{equation}
That is, the set of parameters minimizing the model's simulation error over the training set is computed, subject to the range constraint $\Theta \in \Xi$ that enforces the physical consistency of the parameters\footnote{The parameters considered must all be positive. In case the nominal values of these parameters are known, they can be constrained in a range around these  values.}. 

\begin{remark} \label{rem:velocity}
	In light of the full observability of  \eqref{eq:mechanical:statespace},  penalizing the output simulation error in the loss \eqref{eq:mechanical:loss} is enough to identify the parameters.
	However, if the full state vector $x$ is measured (i.e., $\dot{q}$ is measurable), one can instead resort the state simulation error, which usually yields a faster convergence of the identification algorithm.
	Note that, even if the velocities  $\dot{q}$ are not measured, they can be numerically reconstructed quite accurately offline, for example using the five-point stencil method. 
\end{remark}

\subsection{Shortcomings} \label{sec:mechanical:problems}
Having formulated the grey-box model based on the Euler-Lagrange equations as well as the algorithm for the identification of its parameters, we can highlight the main issues and challenges  of this approach.

First, despite the simplicity of the setup, the approach requires a very thorough knowledge of the mechanical system and its physical parameters. 
Familiarity with the setup is required, for example, to (\emph{i}) understand how to model the elastic effect introduced by the encoder cable~\eqref{eq:lagrangian:spring} which, as will be discussed in Section~\ref{sec:numerical}, is critical to obtaining an accurate model; (\emph{ii}) obtain a meaningful set $\Xi$ for the system's parameters; and (\emph{iii}) provide a good initial guess for the iterative solver of \eqref{eq:mechanical:identification}, that is often crucial to have convergent identification procedures.
A further caveat is that in order to build a sufficiently accurate model in the entire region of interest, the training data must be informative, i.e. it must explore over the region where we want the model to be accurate, and balanced, i.e. the density of the data should be as uniform as possible. 
Prolonged intervals at the equilibrium point, for example, must be removed to avoid overfitting.
All these insights are necessary to achieve a first-principles model with satisfactory accuracy.

At last, it is worth pointing out that while obtaining \eqref{eq:mechanical:accelerations} from \eqref{eq:eulerlagrange}-\eqref{eq:lagrangian} is certainly feasible for the rotary pendulum considered, the procedure gets significantly more involved  for more complex mechanical systems.
In such cases, this approach may prove unpractical.

\section{Physics-based deep learning model} \label{sec:tustin-net}
In this section the physics-based deep learning model known as Tustin-Net is introduced to address the shortcomings of the first-principles model.

\subsection{The Tustin-Net architecture} \label{sec:tustin-net:intro}
\begin{figure}[t]
    \centering
    \includegraphics[width=\linewidth, clip, trim=5mm 0 5mm 0]{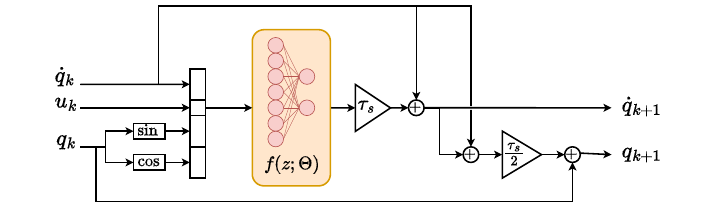}
    \caption{Scheme of the Tustin-Net model.}
    \label{fig:tustin-scheme}
\end{figure}

Tustin-Net is a recently proposed deep learning architecture tailored to the identification of mechanical systems \cite{pozzoli2020tustin}.
This architecture consists in the following discrete-time system
\begin{subnumcases}{\label{eq:tustin:statespace}}
		q_{k+1} = q_k + \tau_s \frac{\dot{q}_k + \dot{q}_{k+1}}{2}, \label{eq:tustin:statespace:1}\\
		\dot{q}_{k+1} = \dot{q}_k + \tau_s f \Big( \text{cat}\big(\sin(q_k), \cos(q_k), \dot{q}_k, u_k \big); \Theta\Big),\label{eq:tustin:statespace:2}
\end{subnumcases}
where \eqref{eq:tustin:statespace:1} represents the known position dynamics, c.f. \eqref{eq:mechanical:accelerations:1}, discretized via the trapezoidal (Tustin) method, whereas  \eqref{eq:tustin:statespace:2} resembles a forward, explicit discretization of \eqref{eq:mechanical:accelerations:2} via the feedforward neural network $f(\cdot; \Theta)$ parametrized by the learnable weights $\Theta$. 
The scheme of the resulting architecture is depicted in Figure~\ref{fig:tustin-scheme}.

This feedforward neural network, which takes as input the sine and cosine features of angular positions, as well as the velocities and the input vector, can be compactly defined as a sequence of nonlinear transformations $\phi_{1}, ..., \phi_{M}$. 
Each of these transformations $ m \in \{ 1, ..., M \}$ is defined as an affine combination of its input by the learnable weights $\Theta_m = \{ W_m, b_m \}$ followed by a Lipschitz-continuous nonlinear activation function $\sigma$, 
\begin{subequations} \label{eq:tustin:ffnn}
\begin{equation}
	\phi_m(z; \Theta_m) = \sigma(W_m z + b_m).
\end{equation}
The neural network thus reads
\begin{equation}
	f(z; \Theta) = W_{M+1} ( \phi_{M} \circ ... \circ \phi_1)(z),
\end{equation}
where $W_{M+1}$ weights the output of the last layer. 
Letting for notational convenience $\Theta_{M+1} = \{ W_{M+1} \}$, the set of learnable weights of the network is
\begin{equation}
	\Theta = \{ \Theta_{1}, ..., \Theta_{M},\Theta_{M+1} \}.
\end{equation}
\begin{remark} \label{rem:neurons}
	The weights $W_m$ and $b_m$  have dimensions $W_m \in \mathbb{R}^{n_{m} \times n_{m-1}}$ and $b_m \in \mathbb{R}^{n_{m}}$, where for dimensional consistency $n_{0} = n_u + 3n_q$ and $n_{M+1} = n_q$. Note that $n_m$, for $m \in \{1, ..., M\}$, is the number of neurons of hidden layers, which are hyperparameters that need to be selected by the user.
\end{remark}
\end{subequations}
In what follows, we compactly denote the Tustin-net model \eqref{eq:tustin:statespace}-\eqref{eq:tustin:ffnn} as
\begin{equation} \label{eq:tustin:compact}
	\begin{dcases}
		x_{k+1} = \psi(x_k, u_k; \Theta), \\
		y_k = \begin{bmatrix}
			I & 0
		\end{bmatrix} x_k.
	\end{dcases}
\end{equation}

\subsubsection{Standard training procedure}
Before describing the adopted transfer learning procedure, let us describe the Tustin-Net training procedure proposed in~\cite{pozzoli2020tustin}.
This training strategy is similar to the identification procedure described in Section~\ref{sec:mechanical:training}, as it consists in iteratively minimizing the loss function \eqref{eq:mechanical:loss} on the training dataset \eqref{eq:mechanical:dataset}.
It is, however, not necessary to impose any constraint on the network weights, as they have no physical meaning.
Peculiarly,  \cite{pozzoli2020tustin} adopts a distance measure that allows the Tustin-Net learning to account for the periodicity of angular coordinates, 
\begin{equation} \label{eq:tustin:distance}
    \begin{aligned}
    \ell_j( [\hat{y}_k ]_j, [y_k ]_j ) &= 
        \big(\cos([\hat{y}_k ]_j) - \cos([{y}_k ]_j) \big)^2 + \big(\sin([\hat{y}_k ]_j) - \sin([{y}_k ]_j) \big)^2 \\
            &= 2 \big(1 - \cos([\hat{y}_k ]_j - [{y}_k ]_j) \big).
    \end{aligned}
\end{equation}

\begin{remark}
    For simplicity of notation, the Tustin-Net formulation \eqref{eq:tustin:statespace} assumes that all the coordinates of $q$ are angular positions. 
    The architecture can be easily extended to cases where $q$ contains both linear and angular positions simply by calculating the sine and cosine features , c.f. \eqref{eq:tustin:statespace:2}, only for the angular positions. In that case, in the loss function the distance measure \eqref{eq:tustin:distance} should only be used for angular positions, while  \eqref{eq:mechanical:distance} should be used for linear positions.
\end{remark}

\subsection{Improving training with transfer learning}
We now propose an alternative transfer learning-based training procedure, which stems from the following observations.

First, we point out that in the domain of nonlinear identification, it has recently been observed that adopting a truncated simulation error minimization yields smoother and more effective training \cite{ribeiro2020smoothness}. 
Indeed, simulating the model over shorter horizons generally enables an enhanced learning of the system dynamics by limiting vanishing/exploding gradient problems \cite{beintema2021nonlinear} and error accumulation.
The training procedure considered here is therefore built on the idea of fitting the model to partially overlapping subsequences extracted from the training data \cite{beintema2021nonlinear, bonassi2022recurrent}.

Second, we propose to fit the entire state vector instead of just the positions.
Although, as discussed in Remark~\ref{rem:velocity}, minimizing the difference between the model-estimated  and the measured positions is sufficient to obtain an accurate model, we observed that exploiting also the measured/estimated velocities allows us to achieve consistently more accurate models that can better generalize to new data.
If, as in the considered setup, the velocities are not measured, they can be effectively estimated offline from the position measurement using, for example, the \emph{five-point stencil} formula
\begin{equation} \label{eq:tustin:velocity_estim}
    [\hat{\dot{q}}_k]_j \approx \frac{[{q}_{k-2}]_j - 8[{q}_{k-1}]_j + 8[{q}_{k+1}]_j - [{q}_{k+2}]_j}{12}.
\end{equation}

Lastly, the performance of the Tustin-Net architecture may suffer from a training set not being properly balanced---for example, because it contains long intervals where the system is at equilibrium.
In such a case, the network is indeed likely to overfit the equilibrium point, thus exhibiting unsatisfactory performance in the entire region of interest.
In line with the \emph{transfer learning} strategy \cite{tan2018survey}, we propose here to (\emph{i}) pre-train the Tustin-Net model on a part of the training data where the system is in transient condition, (\emph{ii}) freeze the weights related to the first layers of the neural network, i.e. $\Theta_m$ for $m \in \{ 1, ..., \tilde{M} \}$ with $\tilde{M} \leq M$, and finally (\emph{iii})  fine-tune the weights of the last layers, i.e. $\Theta_m$ for $m \in \{ \tilde{M}+1, ..., M \}$, on the entire training dataset.
As will be shown in Section~\ref{sec:numerical}, this strategy makes it possible to improve the performance of the resulting model over a wide operating region while mitigating the risks of overfitting.
In the following, these three steps are described in detail.

\subsubsection{Step 1. Network pre-training}
The Tustin-Net \eqref{eq:tustin:compact} is first pre-trained on the dataset $\mathcal{D}_{(1)}$, which is extracted from the training sequences collected in $\mathcal{D}$. The goal of this training stage is to pre-train the network so that it learns the nonlinearities that characterize the state space region of interest, while limiting the risk of the network overfitting the steady-state conditions.
In particular, this dataset is a collection of $N_{(1)} \gg N$ subsequences of length $T_{(1)}$,
\begin{subequations} \label{eq:transfer_learning:D1}
\begin{equation}
	\mathcal{D}_{(1)} = \left\{ (x_0\tridx{l}, u\tridx{l}_{0:T_{(1)}}, x_{0:T_{(1)}}\tridx{l}), \forall l \in \{1, ..., N_{(1)} \}\right\}.
\end{equation}
Each sample $l \in \{1, ..., N_{(1)} \}$ of this set is extracted from the randomly-selected sequence $i_l \sim \mathcal{U}(1, N)$ sequence of $\mathcal{D}$, starting from the step $k_l \sim \mathcal{U}(2, \bar{k}_{i_l}-2)$, randomly drawn before the system is excessively close to an equilibrium point, which time is denoted by $\bar{k}_{i_l}$. Hence 
\begin{equation}
	\begin{aligned}
		x_0\tridx{l} = x_{k_l}\tridx{i_l}, \qquad u\tridx{l}_{0:T_{(1)}} = u_{k_l:k_l+T_{(1)}}\tridx{i_l}, \qquad		x_{0:T_{(1)}}\tridx{l} = x_{k_l:k_l+T_{(1)}}\tridx{i_l}
	\end{aligned}
\end{equation}
\end{subequations}
where, with a slight abuse of notation, the velocity components of  $x_{k_l:k_l+T_{(1)}}\tridx{i_l}$ are intended to be estimated via \eqref{eq:tustin:velocity_estim}.
The pre-training procedure is then carried out by iteratively minimizing the loss function $\mathcal{L}_x(\mathcal{D}_{(1)}; \Theta)$ with respect to the trainable weights, $\Theta$.
Here, the loss function is computed on the state vector, i.e.
\begin{equation} \label{eq:tustin:state_loss}
    \mathcal{L}_x(\mathcal{D}; \Theta) = \frac{1}{\lvert \mathcal{D} \lvert} \sum_{i\in \mathcal{D}} \frac{1}{T_{(1)}} \sum_{k=1}^{T_{(1)}} \sum_{j=1}^{n_y} \ell_j\Big(\big[x_k(x_0\tridx{i}, u_{0:k}\tridx{i}; \Theta)\big]_j, \big[x_{k}\tridx{i}\big]_j \Big),
\end{equation}
where $x_k(x_0\tridx{i}, u_{0:k}\tridx{i}; \Theta)$ is the state of \eqref{eq:tustin:statespace}, initialized in $x_0\tridx{i}$ and fed by the input sequence $u_{0:k}\tridx{i}$.
Here $\ell_j$ is defined as \eqref{eq:tustin:distance} for the angular position state components, i.e. $j \in \{ 1, ..., n_q \}$, and as \eqref{eq:mechanical:distance} for the velocity state components, i.e. $j \in \{ n_q + 1, ..., 2 n_q \}$.

\subsubsection{Step 2. Weights freezing} 
Once the training described in the previous point ends, the weights related to the first $\tilde{M}$ layers are frozen. 
That is, 
\begin{equation} \label{eq:transfer_learning:freeze}
    \Theta_m \equiv \Theta_{m,(1)}^\star \qquad \forall m \in \{1, ..., \tilde{M} \}.
\end{equation}

\subsubsection{Step 3. Fine-tuning}
The Tustin-Net is eventually fine-tuned on a dataset consisting of $N_{(2)}$ subsequences of length $T_{(2)}$, which reads
\begin{subequations} \label{eq:transfer_learning:D2}
\begin{equation}
	\mathcal{D}_{(2)} = \left\{ (x_0\tridx{l}, u\tridx{l}_{0:T_{(2)}}, x_{0:T_{(2)}}\tridx{l}), \forall l \in \{1, ..., N_{(2)} \}\right\}.
\end{equation}
Each sample $l \in \{1, ..., N_{(2)} \}$ is defined by randomly sampling the training sequence index $i_l \sim \mathcal{U}(1, N)$ from which the subsequence is extracted, as well as the initial time-step $k_l \sim \mathcal{U}(2, T_{i_l} - T_{(2)} -2)$,
\begin{equation}
	\begin{aligned}
		x_0\tridx{l} = x_{k_l}\tridx{i_l}, \qquad u\tridx{l}_{0:T_{(2)}} = u_{k_l:k_l+T_{(2)}}\tridx{i_l}, \qquad		x_{0:T_{(2)}}\tridx{l} = x_{k_l:k_l+T_{(2)}}\tridx{i_l}.
	\end{aligned}
\end{equation}
Note that since $T_{i_l}$ denotes the total length of the training sequence $i_l$, the sampled sub-sequences are no longer limited to the initial transient, but can also be sampled from steady-state conditions.

\end{subequations}
The Tustin-Net is thus fine-tuned by iteratively minimizing the loss function $\mathcal{L}_x(\mathcal{D}_{(2)}; \tilde{\Theta})$, c.f. \eqref{eq:tustin:state_loss}, with respect to the weights $\tilde{\Theta} = \{ \Theta_{\tilde{M}+1}, ..., \Theta_{M} \} $. The set of trained weights is finally given by 
\begin{equation} \label{eq:transfer_learning:weights}
    \Theta^\star = \big\{ \Theta^\star_{1,(1)}, ..., \Theta^\star_{\tilde{M},(1)}, \Theta^\star_{\tilde{M}+1,(2)}, ..., \Theta^\star_{M+1,(2)} \big\}.
\end{equation}
The resulting training procedure is summarized in Algorithm~\ref{alg:transfer_learning}.

\begin{algorithm}
\caption{Transfer learning-based training of the Tustin-Net model}
\label{alg:transfer_learning}
\textbf{Input}:  Dataset $\mathcal{D}$ \\
\textbf{Parameters}: $N_{(1)}$, $T_{(1)}$, $N_{(2)}$, $T_{(2)}$, $\tilde{M}$ \\
\textbf{Output}: Trained weights $\Theta^\star$
\begin{algorithmic}
\State Build the dataset $\mathcal{D}_{(1)}$ via \eqref{eq:transfer_learning:D1}
\State Pre-train \eqref{eq:tustin:compact} by iteratively minimizing $\mathcal{L}_x(\mathcal{D}_{(1)}; \Theta)$, yielding $\Theta^\star_{(1)}$
\State Freeze the weights of the first $\tilde{M}$ layers, \eqref{eq:transfer_learning:freeze}
\State Build the dataset $\mathcal{D}_{(2)}$ via \eqref{eq:transfer_learning:D2}
\State Fine-tune \eqref{eq:tustin:compact}  by iteratively minimizing $\mathcal{L}_x(\mathcal{D}_{(2)}; \tilde{\Theta})$, yielding $\tilde{\Theta}^\star_{(2)}$
\State Return the optimal weights $\Theta^\star$ defined as in \eqref{eq:transfer_learning:weights}
\end{algorithmic}
\end{algorithm}

\section{Numerical results and discussion}\label{sec:numerical}
We now report and discuss the performance achieved by the models described in Section~\ref{sec:mechanical} and Section~\ref{sec:tustin-net}, aiming to illustrate the advantages and limitations of all these strategies.

These models are trained on a dataset  of $10$ free-fall experiments, where the pendulum is let fall from an initial state close to the unstable equilibrium ($\alpha_0 \approx 0$), plus $5$ experiments where the system is initialized near the stable equilibrium ($\alpha_0 \approx \pi$) and a white noise voltage is applied to the motor. 
The dataset \eqref{eq:mechanical:dataset} thus consists of $N=15$ input-output sequences, sampled with sampling time $\tau_s = 10^{-2}~\textrm{s}$ and with a length ranging from $7.16$  to $14.99$ seconds.
The data, along with code to train Tustin-Nets, is available in the accompanying Github repository\footnote{\url{https://github.com/svanesch/tustinNetTransferLearning}}.

In order to compare the performance of the identified models, $8$ additional experiments have been conducted on the pendulum ($4$ free-fall experiments and $4$ experiments with white-noise excitation input), which constitute the independent \emph{validation dataset} on which the performance metrics of the models will be evaluated.
We here consider as performance indicator the free-run simulation's Root Mean Squared Error (RMSE). That is,
\begin{equation}
    \textrm{RMSE} \left(y_{0:T_{\text{v}}}, y_{0:T_{\text{v}}}^{\{\text{v}\}} \right) = \sqrt{\frac{1}{T_{\text{v}}} \sum_{k=0}^{T_{\text{v}}} \| y_k - y_k^{\{\text{v} \}} \|_2^2},
\end{equation}
where $\big(x_{0}\tridx{\text{v}}, u_{0:T_{\text{v}}}\tridx{\text{v}},  y_{0:T_{\text{v}}}\tridx{\text{v}} \big)$ denotes the data of a generic validation experiment, and $y_{0:T_{\text{v}}} = y_{0:T_{\text{v}}}\big(x_{0}\tridx{\text{v}}, u_{0:T_{\text{v}}}\tridx{\text{v}}; \Theta^\star \big)$ denotes the free-run simulation of the model under consideration.

\subsection{Euler-Lagrange model}
In Section~\ref{sec:mechanical}, we mentioned that the encoder cable introduces a nonlinear force on the base arm, and that modeling such a force by  including a suitable elastic term $V_s(q, \Theta)$ in the potential energy \eqref{eq:lagrangian:potential} is paramount to achieve a good accuracy of the Euler-Lagrange model.

\begin{figure}[t]
    \centering
    \includegraphics[width=0.95\linewidth, clip, trim=0 4mm 0 0]{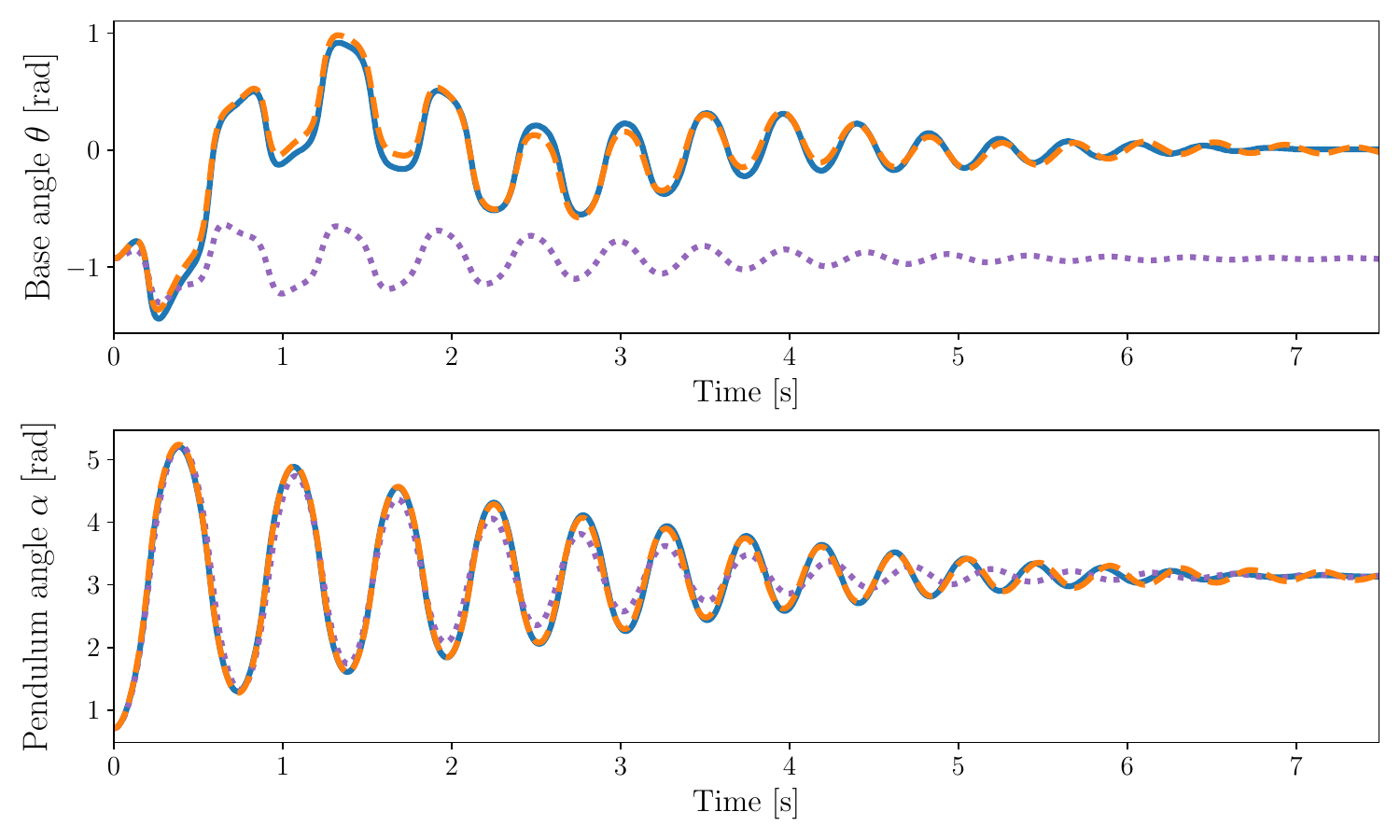}
    \caption{Free-run simulation of the Euler-Lagrange model \eqref{eq:mechanical:statespace} with (orange dashed line) and without (purple dotted line) spring model, compared the ground truth (blue continuous line) on one of the validation experiments.}
    \label{fig:freefall:spring}
\end{figure}

To support this claim, in Figure~\ref{fig:freefall:spring} we compare the free-run simulation of the Euler-Lagrange model \eqref{eq:mechanical:statespace}, which captures the effect of the cable via the polynomial elastic model \eqref{eq:lagrangian:spring}, to that of a na\"ive Euler-Lagrange model without this term, i.e. with $V_s(q; \Theta) = 0$. 
Despite them both being identified via \eqref{eq:mechanical:identification} with the same training dataset, it is evident that not modeling the elastic effect results in significantly worse performance ($\text{RMSE} = 0.966$), as the model fails to capture the nonlinearity of the base arm dynamics, particularly observable in the angle~$\theta$. 
On the other hand, when the elastic term is included, the identified Euler-Lagrange model achieves remarkable modeling performance ($\text{RMSE} = 6.7 \times 10^{-2}$).
This results emphasize that while first-principles models can describe the system's dynamics quite accurately, a comprehensive understanding of the setup is often necessary to achieve satisfactory modeling performance.

\subsection{Tustin-Net models}
The Tustin-Net here considered consists of $M=2$ hidden layers with $n_{1} = n_2 = 100$ units, see Remark~\ref{rem:neurons}. 
The Leaky ReLU activation function \cite{goodfellow2016deep} has been adopted.
The network's sampling time hyperparameter has been fixed to $\tau_s = 0.01$ s, which matches the sampling time of the training data.

The Tustin-Net has been trained with the transfer learning algorithm summarized in Algorithm~\ref{alg:transfer_learning}.
In particular, in the pre-training stage the Tustin-Net has been trained on $N_{(1)} = 1408$ sub-sequences of length $T_{(1)} = 50$ extracted from the training dataset as described in \eqref{eq:transfer_learning:D1}.
After 300 training epochs, the weights of both hidden layers have been frozen, i.e. $\tilde{M} = M = 2$, while the output layer's weights $\Theta_{M+1}$ are fine-tuned in the last stage on $N_{(2)} = 864$ sub-sequences of length $T_{(2)} = 75$ extracted from the entire training sequences, c.f. \eqref{eq:transfer_learning:D2}. 
During the training procedures, a learning rate scheduler has been implemented to reduce the learning rate in case of plateaus.

In order to assess the impact of the transfer learning procedure, we compare the modeling performance of the Tustin-Net trained by Algorithm~\ref{alg:transfer_learning} to that of a Tustin-Net (with the same size) trained by the standard procedure described in Section~\ref{sec:tustin-net:intro}.\footnote{Because of numerical instabilities and computational bottlenecks in the standard training procedure, also for this training  the sequences have been divided into shorter, partially overlapping subsequences.}
Note that, to ensure a fair comparison, this latter has been trained for a number of epochs such that the number of processed datapoints matches the total number of datapoints processed at all the stages of Algorithm~\ref{alg:transfer_learning}.

\begin{figure}[t]
    \centering
    \includegraphics[width=0.95\linewidth, clip, trim=0 4mm 0 0]{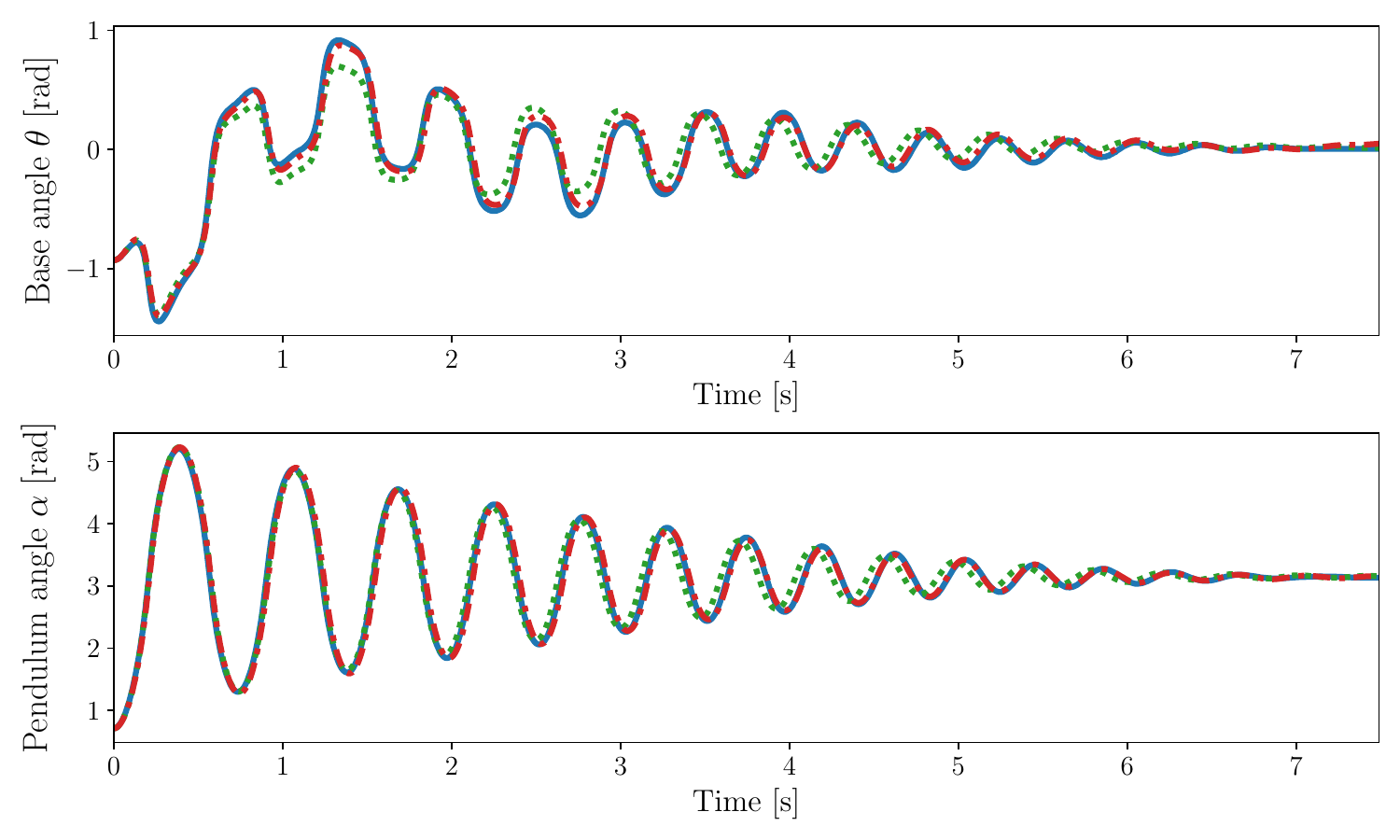}
    \caption{Free-run simulation of the Tustin-Net model \eqref{eq:tustin:statespace} learned via Algorithm~\ref{alg:transfer_learning} (red dashed-dotted line) compared to the ground truth (blue continuous line) and to standard training procedure (Section \ref{sec:tustin-net:intro}) on a validation dataset.}
    \label{fig:freefall:transferLearning}
\end{figure}

The free-run simulation of these two networks, reported in Figure~\ref{fig:freefall:transferLearning}, witnesses how the transfer learning procedure allows the Tustin-Net to learn both the nonlinearities in the initial transient and the almost-linear dynamics around the equilibrium.

\begin{table}[b]
    \centering
    \caption{Performance comparison on validation data -- \textrm{RMSE} $[\times 10^{-1}]$}
    \label{tab:mse}
    \resizebox{\columnwidth}{!}{
    \begin{tabular}{lcccccccccc}
    \toprule
    \multirow{2}{*}{\quad} & \qquad &
      \multicolumn{4}{c}{Free-fall experiments} & \qquad &
      \multicolumn{4}{c}{White noise excitation} \\
    Model &\quad\qquad& \#1 & \#2 & \#3 & \#4 &\quad\qquad& \#5 & \#6 & \#7 & \#8  \\
    \midrule
    Euler-Lagrange  && \, 1.42 \, & \, {1.07} \, & \, 3.11 \, & \, {0.67} \, && \, 1.41 \,& \, 1.61 \, & \, 1.39 \, & \,  2.99 \,\\
    Tustin-Net (standard training) && \, 5.31 \, & \, 2.28 \, & \, 6.14 \, & \, 1.84 \, && \, 0.72 \, & \, 1.23 \, & \, 1.57 \, & \, 1.79 \,\\
    Tustin-Net (Algorithm~\ref{alg:transfer_learning}) && \, 0.75 \, & \, 1.53 \, & \, 1.11 \, & \, 0.94 \, && \, 1.03 \, & \, 1.42 \, & \, 1.68 \, & \, 1.77 \,\\
    \bottomrule
    \end{tabular}}
\end{table}

\subsection{Summary and discussion}
\begin{figure}[t]
    \centering
    \includegraphics[width=0.95\linewidth, clip, trim=0 4mm 0 0]{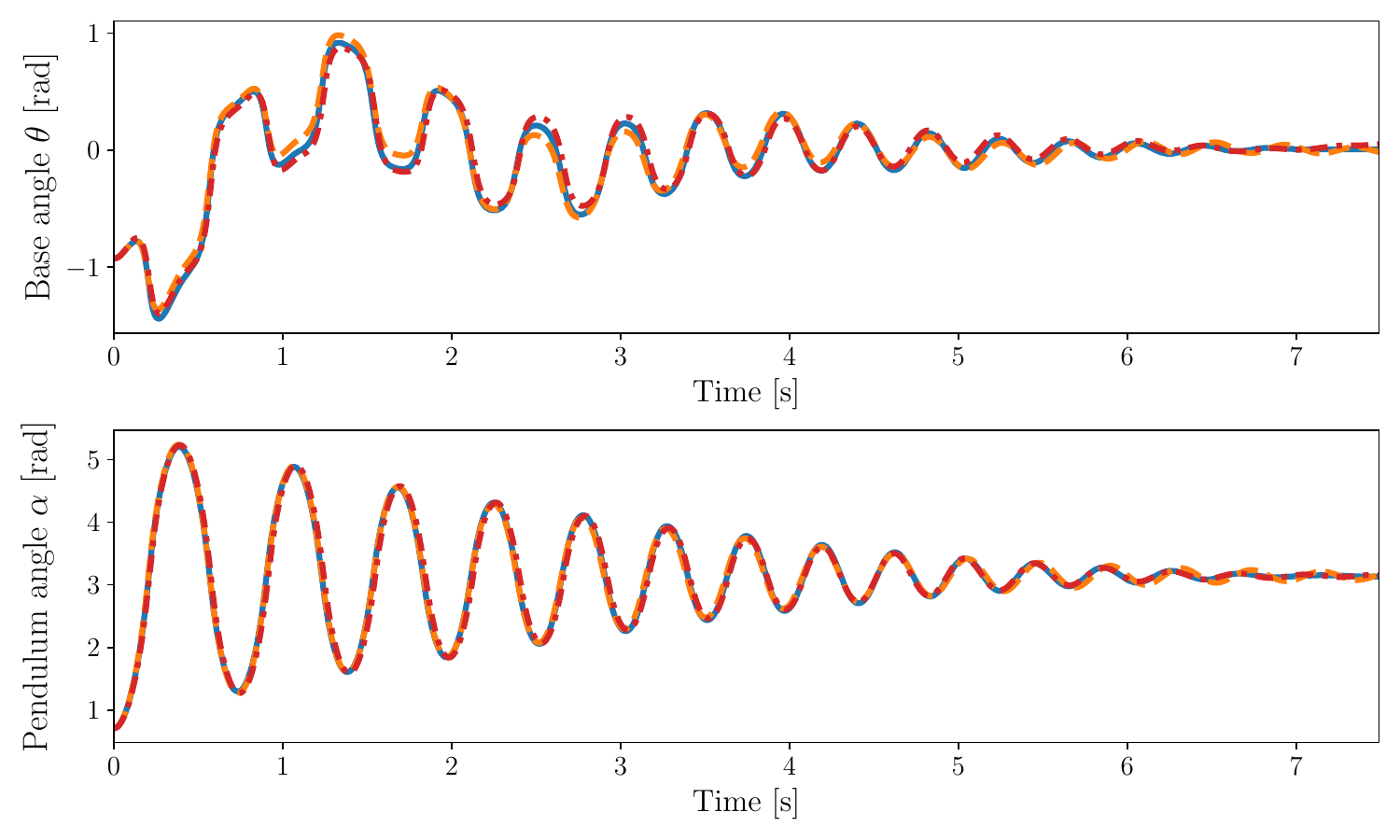}
    \caption{Performance comparison of the Tustin-Net model learned by Algorithm~\ref{alg:transfer_learning} (red dashed-dotted line) and Euler-Lagrange model (orange dashed line) on the validation dataset.}
    \label{fig:freefall:ElvsTustin}
\end{figure}

Let us now compare the performance of Tustin-Nets learned through Algorithm~\ref{alg:transfer_learning} with the Euler-Lagrange model.
In Figure~\ref{fig:freefall:ElvsTustin} the free run simulations on the same test experiment are directly compared. As can be seen, the models have successfully learned the system dynamics and thus achieve similar accuracy.

To provide a more comprehensive comparison, in Table~\ref{tab:mse} the free-run simulation RMSE on all test experiments are reported.
It can be noticed that the Tustin-Net trained via transfer learning enjoys consistently better performance than that of the network trained by the standard procedure, thus confirming the above-described advantages of the proposed training method.

The results obtained from the Tustin-Nets are in line with those of the physical model based on the Euler-Lagrange equations.
We can however note that, unlike first-principles models, Tustin-Nets only require very little knowledge of the physical setup, a principled data collection campaign, and a proper selection of the model's and training's hyperparameter.
On the other hand, the Euler-Lagrange model requires a thorough knowledge of the physical system to be modeled, the physical laws that govern it, and the non-idealities that may characterize it.
While this modeling effort may be time consuming, especially for more complex system, first-principles models are known to generalize better to operational regions not explored in the training dataset.

\section{Conclusions} \label{sec:conclusions} 
In this report we discussed several strategies for modeling a rotational pendulum, ranging from a grey-box model based on the Euler-Lagrange equations to a physics-based deep learning model known as Tustin-Net.
Training this architecture involves the learning of the discretized velocity dynamics with a feedforward neural network, while position dynamics are modeled via trapezoidal integration.

One of the problems that can arise in training Tustin-Nets is that if the dataset is not balanced, the network can overfit operating conditions and fail to accurately approximate nonlinear transients.
We therefore proposed a training procedure, based on transfer learning paradigm, that builds on the idea of pre-training a Tustin-Net model on the initial transients of the training data and then fine-tuning the weights of the last layers of the network on the entire dataset.

The proposed learning strategy has proven to yield Tustin-Net models with comparable, if not better, performance than first-principles models, with the advantages of not requiring physical knowledge of the system to be identified and being less sensitive to imbalance in the training dataset.

\bibliography{biblio}

\end{document}